\documentclass[prd,nofootinbib,preprint,superscriptaddress]{revtex4-2}

\usepackage{amsmath, amssymb, amsthm, graphicx, epsfig, fancyhdr,epsfig, slashed}

\usepackage{tikzsymbols}
\usepackage{natbib}
\usepackage{float}

\usepackage{tikz,xcolor,hyperref}

\definecolor{lime}{HTML}{A6CE39}
\DeclareRobustCommand{\orcidicon}{
	\begin{tikzpicture}
	\draw[lime, fill=lime] (0,0) 
	circle [radius=0.2] 
	node[white] {{\fontfamily{qag}\selectfont \tiny ID}};
	\draw[white, fill=white] (-0.0625,0.095) 
	circle [radius=0.007];
	\end{tikzpicture}
	\hspace{-2mm}
}

\foreach \x in {A, ..., Z}{\expandafter\xdef\csname orcid\x\endcsname{\noexpand\href{https://orcid.org/\csname orcidauthor\x\endcsname}
			{\noexpand\orcidicon}}
}

\newcommand{\be}{\begin{equation}}
\newcommand{\ee}{\end{equation}}
\newcommand{\bea}{\begin{eqnarray}}
\newcommand{\eea}{\end{eqnarray}}

\newcommand{\ag}[1]{\textcolor{blue}{[Anish: #1]}}

\begin{document}

\title{Non-perturbative Origin of Electroweak Scale via Higgs-portal:\\  \it{Dyson-Schwinger in Conformally Invariant Scalar Sector} }

\author{Marco Frasca\orcidA{}}
\email{marcofrasca@mclink.it}
\affiliation{Rome, Italy}

\author{Anish Ghoshal\orcidB{}}
\email{anish.ghoshal@fuw.edu.pl}
\affiliation{Institute of Theoretical Physics, Faculty of Physics, University of Warsaw, ul. Pasteura 5, 02-093 Warsaw, Poland}

\author{Nobuchika Okada\orcidC{}}
\email{okadan@ua.edu}
\affiliation{Department of Physics and Astronomy, \\ University of Alabama, Tuscaloosa, AL 35487, USA}

\begin{abstract}
\textit{We investigate conformally extended Standard Model with a hidden scalar $\phi$. It is shown that due to non-perturbative dynamics in the hidden sector, $\phi$ develops a vacuum expectation value (vev) in the form of a mass gap which triggers the electroweak symmetry breaking (EWSB) and dynamically generates the SM Higgs boson mass. For estimating the non-perturbatively generated mass scale, we solve the hierarchy of Dyson-Schwinger Equations in form of partial differential equations using the exact solution known via a novel technique developed by Bender, Milton and Savage. We employ Jacobi Elliptic function as exact background solution and show that the mass gap that arises in the hidden sector can be transmuted to the EW sector, expressed in terms of Higgs-portal mixed quartic coupling $\beta$ and self interaction quartic coupling $\lambda_{\phi}$ of $\phi$. We identify the suitable parameter space where the observed SM Higgs boson can be successfully generated . Finally, we discuss how this idea of non-perturbative EW scale generation can serve as a new starting point for better realistic model building in the context of resolving the hierarchy problem in the Standard Model.} 
\end{abstract}

\maketitle

\section{Introduction}

The presence of any fundamental scalar field in quantum field theory (QFT) encounters the hierarchy problem because of the fact that its mass should be of the same order as that of the cut-off scale of the theory due to quantum corrections. As an example, if Planck scale is the highest scale in the theory, the Standard Model (SM) Higgs, if it is a fundamental scalar, should receive quantum radiative corrections to its mass of order of Planck scale, where quantum gravity is thought to be dominant \footnote{However, recently in higher-derivative non-local extensions of QFT, inspired by p-adic string field theories, it has been shown that this problem can be relaxed, and conformal invariance can be dynamically achieved without introducing any new particles in the physical mass spectrum, see Refs. \cite{Ghoshal:2017egr,Ghoshal:2018gpq,Ghoshal:2020lfd,Frasca:2020jbe,Frasca:2020ojd,Frasca:2022duz} with very concrete predictions and interesting LHC phenomenology \cite{Biswas:2014yia,Su:2021qvm}.}.  

Among several possibilities including supersymmetry and extra-dimensional theories one popular and quite elegant solution to this problem is to hypothesize \textit{scale invariance} as a symmetry of the fundamental action, such that all scales we observe in Nature be generated dynamically beyond classical level. Coleman and Weinberg \cite{Coleman:1973jx} 
proposed such a scenario and showed that the SM gauge symmetry
breaking could be radiatively triggered via quantum corrections. However, from the observational point-of-view this mechanism fails within the Standard Model to generate the correct Higgs mass (the Electroweak (EW) Scale) as it predicts
$m_{Z,W} > m_H$, where $m_{Z,W}$ are the SM Z and W gauge boson masses while $m_H$ is the SM Higgs boson mass \cite{Coleman:1973jx,Englert:2013gz}. In spite of this failure this has remained the direction of BSM model building and several extensions of the SM has been explored extensively in the literature \cite{Adler:1982ri,Coleman:1973jx,Salvio:2014soa,Einhorn:2014gfa,Einhorn:2016mws,Einhorn:2015lzy}, where this radiative EW symmetry breaking (EWSB) mechanism is successful in terms of observations and often makes very concrete and testable predictions. When non-minimal coupling to gravity is introduced, such scenarios can provide naturally flat inflaton potentials \cite{Khoze:2013uia,Kannike:2014mia,Rinaldi:2014gha,Salvio:2014soa,Kannike:2015apa,Kannike:2015fom,Barrie:2016rnv,Tambalo:2016eqr} and stable particle dark matter candidates \cite{Hambye:2013sna,Karam:2015jta,Kannike:2015apa,Kannike:2016bny,Karam:2016rsz,Barman:2021lot}. 
The scenarios also lead to very strong first-order phase transitions and hence the possibility of high amplitude detectable gravitational wave (GW) signals in upcoming detectors~\cite{Jaeckel:2016jlh, Marzola:2017jzl, Iso:2017uuu, Baldes:2018emh, Prokopec:2018tnq, Brdar:2018num, Marzo:2018nov, Ghoshal:2020vud}. 
Consequently, scale invariant scenarios offer an interesting direction of model-building for solving the hierarchy problem in the Standard Model of particle physics
\cite{Foot:2007iy,AlexanderNunneley:2010nw,Englert:2013gz,Hambye:2013sna,Farzinnia:2013pga,Altmannshofer:2014vra,Holthausen:2013ota,Salvio:2014soa,Einhorn:2014gfa,Kannike:2015apa,Farzinnia:2015fka,Kannike:2016bny}. 
See Refs. \cite{1306.2329,1410.1817,0902.4050,0909.0128,1210.2848,1703.10924,1807.11490} for other studies of conformal invariance and dimensional transmutation of energy scales along similar lines \cite{2012.11608,1812.01441,2201.12267,2011.10586,2010.15919,2102.10665,1812.02314,1709.09222}.

Note that all the above mentioned analysis considered only the weak perturbation theory. In this paper, we develop a novel method to investigate such dynamical Higgs mass generation due to non-perturbative dynamics. We particularly focus on an scalar extension of the SM with a hidden sector dynamics triggering the EWSB. We develop a novel technique to solve series of Dyson-Schwinger Equations using exact Green's function solution of the background Equation of Motion involving Jacobi Elliptical function, following the analytic approach of Dyson-Schwinger equations which is originally devised by Bender, Milton and Savage in Ref.~\cite{Bender:1999ek}. Since the approach remains valid even in the strongly-coupled regime \cite{Frasca:2015yva}, this technique has been recently applied to QCD in Refs.~\cite{Frasca:2021yuu,Frasca:2021mhi,Frasca:2022lwp,Frasca:2022pjf,Chaichian:2018cyv} and to the SM Higgs sector in Ref.~\cite{Frasca:2015wva}, as well as to other types of SM extensions over the past two decades, see Refs.~\cite{Frasca:2019ysi, Chaichian:2018cyv, Frasca:2017slg, Frasca:2016sky, Frasca:2015yva, Frasca:2015wva, Frasca:2013tma, Frasca:2012ne, Frasca:2009bc, Frasca:2010ce, Frasca:2008tg, Frasca:2009yp, Frasca:2008zp, Frasca:2007uz, Frasca:2006yx, Frasca:2005sx, Frasca:2005mv, Frasca:2005fs}. Recently, these have been employed to study non-perturbative hadronic contributions to the muon anomalous magnetic moment (g-2)$_{\mu}$ \cite{Frasca:2021yuu}, QCD in the non-perturbative regime \cite{Frasca:2021mhi,Frasca:2022lwp,Frasca:2022pjf}, non-perturbative false vacuum decay \cite{Frasca:2022kfy}, as well as to explore the mass gap and confinement in string-inspired infinite-derivative and Lee-Wick theories \cite{Frasca:2020jbe,Frasca:2020ojd,Frasca:2021iip}.

In the literature, the very simple idea of dynamical generation of EW scale from strong interactions has been studied extensively \cite{Hill:2002ap,Lane:2002sm,Sannino:2008ha}, a scenario very well-known as techni-color in old days. In its earliest version, the EW interactions of the matter content, in the form of fermions like techni-quarks Q were postulated such that their condensates breaks the EW symmetry and the EW scale originates from the dynamical scale of the technicolor physics. Later on this scenario was found disfavored in term of the flavor observables, the EW precision data and the measurements of the Higgs properties after its discovery.
Later on, alternative strong dynamics was invoked to generate a composite or partially-composite Higgs boson, which more or less had an approach of postulating effective Lagrangians than fully realistically complete models. Phenomenologically viable dynamical models leading to interesting LHC phenomenology have been studied in \cite{Kilic:2009mi,Klett:2022iga}, although these models do not break EW symmetry nor provide a composite Higgs.

Assuming that quadratically divergent corrections to the Higgs mass squared have no physical meaning and hence can be ignored, possibly because the fundamental theory does
not contain any mass term, one may promote a scale-invariant\footnote{We will be using ``scale-invariance" and ``conformal invariance" inter-changeably in our paper, since it has been shown that they are classically equivalent for any four-dimensional field theory which respects unitarity and renormalizability\cite{Gross:1970tb,Callan:1970yg,Coleman:1970je}.} symmetry principle at the classical level. In this context, dynamical generation of the EW scale
via dimensional transmutation can be realised in models
where an extra scalar $\phi$ develps a vacuum expectation value (VEV) through non-perturbative from $\lambda|\phi|^4$ interaction and then its interaction $\beta H^2|\phi|^2$ effectively generates a negative Higgs mass squared $m^2\sim -\beta\langle\phi\rangle^2$ with $\beta>0$ for the SM Higgs doublet $H$. 

In order to study the non-perturbative dynamics for $\lambda_\phi\phi^4$ interaction, we utilize the exact solutions found in terms of the Jacobi elliptical functions following the analytic approach of Dyson-Schwinger equations originally devised by Bender, Milton and Savage in Ref.~\cite{Bender:1999ek}. In this case, the Green's functions of the theory are represented analytically, and therefore it is straightforward to understand the effect of the background on the interactions that remain valid even in the strongly-coupled regime \cite{Frasca:2015yva}. 

The paper is organized as follows: we start in the next section by a short review of the strongly-coupled technique to generate mass gap via solving Dyson-Schwinger Equations in terms of Jacobi Elliptic function. In section III, we will discuss a simple extension of the SM involving Higgs-portal extra scalar singlet $\phi$ and generate EWSB dynamically. Section IV is devoted to conclusions and discussions.

\medskip

\section{Mass Gap via Jacobi Elliptic Functions: short review}



As a starting point, one considers the partition function. E.g., for a theory with action $S[\phi]$ one has
\begin{equation}
Z[j]={\cal N}\int [d\phi]e^{iS[\phi]-i\int d^4xj(x)\phi(x)}
\end{equation}
with a scalar field $\phi(x)$. It is obvious that this functional integral does not change after a re-parametrization $\phi(x)\rightarrow\phi(x)+\alpha(x)$, with an arbitrary function $\alpha(x)$. Therefore,
\begin{equation}
Z[j]\rightarrow Z[j]\langle e^{i\int d^4x\alpha(x)\left(\frac{\delta S}{\delta\phi(x)}-j(x)\phi(x)\right)}\rangle_j,
\end{equation}
from which, by requiring invariance, we derive the quantum equation of motion:
\begin{equation}
\label{eq:dsder}
\left\langle\frac{\delta S}{\delta\phi(x)}\right\rangle_j=j(x).
\end{equation}
Repeating the derivation with respect to the current $j$
will give all the full set of Dyson-Schwinger equations for the correlation functions after setting $j=0$ at the end of computation. We note that from the lhs of eq.~(\ref{eq:dsder}) one gets the average on the classical equations of motion of the theory that are the starting point for the procedure.

Bender-Milton-Savage method \cite{Bender:1999ek} takes the move from eq.~(\ref{eq:dsder}), working always with higher-order n-point functions $G_n(x_1,x_2,\ldots,x_n)$, without explicitly introducing vertex parts. This permits to preserve the differential structure of the Dyson-Schwinger equations, making this approach particularly useful when exact solutions are known. E.g., this will give for a $\phi^4$ theory \cite{Frasca:2015yva}
\begin{eqnarray}
\label{eq:G1G2}
&&\partial^2G_1+m^2 G_1+\lambda G_1^3+3\lambda G_2(0)+\lambda G_3(x,x,x)=0, \nonumber \\
&&(\partial^2+m^2)G_2(x-y)+3\lambda G_2(0)G_2(x-y)+3\lambda[G_1(x)]^2G_2(x-y) \nonumber \\ 
&&+3\lambda G_3(x,x,y)G_1(x)+\lambda G_4(x,x,x,y))= \delta^4(x-y).
\end{eqnarray}
Exact and non-trivial solutions for $G_1$ are now known: for $G_3(x,x,x)=0$, the the set of Dyson-Schwinger equations becomes treatable without any truncation.

Next we will move onto an SM extension where this BSM sector will be able to generate EWSB dynamically.

\medskip



\section{SM Higgs mass from hidden sector mass gap}
\label{}

Suppose the system has calssically conformal symmetry. Now the only possibility to write terms in 4-D with 2 scalar fields and strictly re-normalizable (in t'Hooft-Veltman sense) is the following Lagrangian: 




\begin{equation}
\label{eq:mainmod}
    \mathcal{L}=\frac{1}{2}(\partial\phi)^2 +\frac{1}{2}(\partial h)^2
    -\frac{\lambda_\phi}{4}\phi^4-\frac{\lambda_h}{4}h^4
    +\beta\phi^2h^2,
\end{equation}
where $\phi$ is a scalar field, $h$ is the SM Higgs field, $\beta$ is their interaction coupling, $\lambda_\phi$ and $\lambda_h$ are the self-interaction couplings. We assume $\lambda_h\ll 1$ and $\beta\ll\lambda_\phi$.

Before to go through the full model, it would be helpful to understand the decoupled case with $\beta=0$. One is left with a quartic scalar theory. This case has been extensively studied by us in \cite{Frasca:2015yva,Chatterjee:2024dgw} and two kinds of solutions are expected. The quartic model admits an exact solutions in terms of Jacobi elliptical functions for eq.(\ref{eq:G1G2}) and all higher order correlation functions are expressed through them that is the hallmark of a Gaussian solution. On the other hand, as proven in \cite{Chatterjee:2024dgw}, there is also the constant solution that arises from quantum fluctuations as in eq.(\ref{eq:G1G2}) the term $G_2(0)$, when properly evaluated and regularized, yields a mass term. This term has the right sign to give rise to the Higgs mechanism as currently understood for the Standard Model. Thus, it is up to nature to decide what solution applies. For simplicity reasons, the Higgs mechanism appears the most economical one and seems the one observed in experiments. Indeed, the other solution implies an infinite tower of massive excitations that were not observed so far\footnote{A comparison with the Coleman-Weinberg effective potential \cite{Coleman:1973jx} could only be possible in the approximate limit of a small coupling while, in our case, we have an exact solution that holds at any value of the coupling itself.}.

\medskip


\subsection{Dyson-Schwinger equations}

Classical equations of motion are easily obtained to be
\begin{eqnarray}
\partial^2\phi &=& -\lambda_\phi\phi^3+\beta h^2\phi+j_\phi, \nonumber \\
\partial^2 h &=& -\lambda_h h^3+\beta\phi^2 h + j_h.
\end{eqnarray}
Given the partition function
\be
Z[j_\phi,j_h]=\int [d\phi][dh]\exp\left[-\int d^4xL-\int d^4x(j_\phi\phi+j_hh)\right],
\ee
we can evaluate the averages as
\bea
\partial^2 G_1^\phi&=&-\lambda_\phi Z^{-1}[j_\phi,j_h]\langle\phi^3\rangle+\beta Z^{-1}[j_\phi,j_h]\langle h^2\phi\rangle +j_\phi \nonumber \\
\partial^2 G_1^h&=&-\lambda_hZ^{-1}[j_\phi,j_h]\langle h^3\rangle+\beta Z^{-1}[j_\phi,j_h]\langle\phi^2 h\rangle + j_h.
\eea
We notice that
\bea
&&G_1^\phi(x) Z[j_\phi,j_h]=\langle\phi(x)\rangle \nonumber \\
&&G_2^{\phi\phi}(x,x)Z[j_\phi,j_h]+[G_1^\phi(x)]^2Z[j_\phi,j_h]=\langle\phi^2(x)\rangle \nonumber \\
&&G_3^{\phi\phi h}(x,x,x)Z[j_\phi,j_h]+G_2^{\phi\phi}(x,x)G_1^h(x)Z[j_\phi,j_h]+
2G_2^{\phi h}(x,x)G_1^\phi(x)Z[j_\phi,j_h]+ \nonumber \\
&&[G_1^\phi(x)]^2G_1^h(x)Z[j_\phi,j_h]=\langle\phi^2(x)h(x)\rangle
\eea
Similarly, interchanging $\phi$ with $h$ will yield
\bea
&&G_1^h(x) Z[j_\phi,j_h]=\langle h(x)\rangle \nonumber \\
&&G_2^{hh}(x,x)Z[j_\phi,j_h]+[G_1^h(x)]^2Z[j_\phi,j_h]=\langle h^2(x)\rangle \nonumber \\
&&G_3^{hh\phi}(x,x,x)Z[j_\phi,j_h]+G_2^{hh}(x,x)G_1^\phi(x)Z[j_\phi,j_h]+
2G_2^{h\phi}(x,x)G_1^h(x)Z[j_\phi,j_h]+ \nonumber \\
&&[G_1^h(x)]^2G_1^\phi(x)Z[j_\phi,j_h]=\langle h^2(x)\phi(x)\rangle.
\eea
Therefore, one has the equations for the 1P-functions for $\phi$ field
\bea
\label{eq:G1phi}
&&\partial^2 G_1^\phi(x)+\lambda_\phi\left\{[G_1^\phi(x)]^3+3G_2^{\phi\phi}(x,x)G_1^\phi(x)
+G_3^{\phi\phi\phi}(x,x,x)\right\}= \nonumber \\
&&\beta\left\{
G_3^{hh\phi}(x,x,x)+G_2^{hh}(x,x)G_1^\phi(x)+
2G_2^{h\phi}(x,x)G_1^h(x)+
[G_1^h(x)]^2G_1^\phi(x)
\right\}+j_\phi,
\eea
and for the Higgs field
\bea
\label{eq:G1h}
&&\partial^2 G_1^h(x)+\lambda_h\left\{[G_1^h(x)]^3+3G_2^{hh}(x,x)G_1^h(x)
+G_3^{hhh}(x,x,x)\right\}= \nonumber \\
&&\beta\left\{
G_3^{\phi\phi h}(x,x,x)+G_2^{\phi\phi}(x,x)G_1^h(x)+
2G_2^{\phi h}(x,x)G_1^\phi(x)+
[G_1^\phi(x)]^2G_1^h(x)
\right\}+j_h.
\eea
Setting to zero the 3P-function at the same point and the currents, one has for the 1P-functions
\bea
&&\partial^2H_1^\phi(x)+\lambda_\phi\left\{[H_1^\phi(x)]^3+3H_2^{\phi\phi}(0)H_1^\phi(x)\right\}= \nonumber \\
&&\beta\left\{H_2^{hh}(0)H_1^\phi(x)+2H_2^{h\phi}(0)H_1^h(x)+[H_1^h(x)]^2H_1^\phi(x)\right\},
\eea
and for the Higgs field
\bea
\label{eq:H1h}
&&\partial^2H_1^h(x)+\lambda_h\left\{[H_1^h(x)]^3+3H_2^{hh}(0)H_1^h(x)\right\}= \nonumber \\
&&\beta\left\{H_2^{\phi\phi}(0)H_1^h(x)+
2H_2^{\phi h}(0)H_1^\phi(x)+
[H_1^\phi(x)]^2H_1^h(x)
\right\}.
\eea
We further set $H_2^{h\phi}(0)=H_2^{\phi h}(0)=0$ and get the full symmetrical set
\bea
\label{eq:H1phi}
&&\partial^2H_1^\phi(x)+\lambda_\phi[H_1^\phi(x)]^3+3\lambda_\phi H_2^{\phi\phi}(0)H_1^\phi(x) \nonumber \\
&&-\beta H_2^{hh}(0)H_1^\phi(x)-\beta[H_1^h(x)]^2H_1^\phi(x)=0,
\eea
and for the Higgs field
\bea
\label{eq:H1h1}
&&\partial^2H_1^h(x)+\lambda_h[H_1^h(x)]^3+3\lambda_hH_2^{hh}(0)H_1^h(x)\nonumber \\
&&-\beta H_2^{\phi\phi}(0)H_1^h(x)-\beta[H_1^\phi(x)]^2H_1^h(x)=0.
\eea
Now, we can make an approximation that $\beta[H_1^\phi(x)]^2H_1^h(x)$ is small with respect to the other terms in the equation and the solution $H_1^h(x)=v$ with a constant $v$, holds at the leading order (mean field approximation). This will yield for the $\phi$ field
\be
\label{eq:H1_0}
\partial^2H_1^\phi(x)+\lambda_\phi[H_1^\phi(x)]^3
+\mu^2_\phi H_1^\phi(x)=0,
\ee
where
\be
\mu_\phi^2=3\lambda_\phi H_2^{\phi\phi}(0)
-\beta H_2^{hh}(0)-\beta v^2.
\ee
This equation can be solved exactly. Note that this approximation holds even if the $\phi$ field is strongly coupled. For consistency reason, one should have in eq.(\ref{eq:H1h1})
\be
3\lambda_h H_2^{hh}(0)-\beta H_2^{\phi\phi}(0)<0,
\ee
for the Higgs sector. This grants the correct vacuum expectation value for the theory for small $\beta$.

\subsection{Gap equations}

We can solve eq.(\ref{eq:H1_0}) and obtain the corresponding 2P-function of the form:
\begin{equation}
H_1^{\phi}(x)=\sqrt{\frac{2\mu^4}{\mu_\phi^2+\sqrt{\mu_\phi^4+2\lambda_\phi\mu^4}}}{\rm sn}\left(p\cdot x+\chi,\kappa\right),
\end{equation}
where $\mu$ and $\chi$ are arbitrary integration constants, $\kappa=\frac{-\mu_\phi^2+\sqrt{\mu_\phi^4+2\lambda_\phi\mu^4}}{-\mu_\phi^2-\sqrt{\mu_\phi^4+2\lambda_\phi\mu^4}}$, 
and the momentum $p$ is given by
\begin{equation}
\label{eq:disp}
    p^2=\mu_\phi^2+\frac{\lambda_\phi\mu^4}{\mu_\phi^2+\sqrt{\mu_\phi^4+2\lambda_\phi\mu^4}}.
\end{equation}
We need the 2P-functions that can be obtained from eq.(\ref{eq:G1phi}) and (\ref{eq:G1h}). Setting the currents to zero at the end of computation, one has
\bea
\label{eq:G1phi0}
&&\partial^2 H_2^{\phi\phi}(x,y)
+3\lambda_\phi [H_1^\phi(x)]^2H_2^{\phi\phi}(x,y)
+3\lambda_\phi H_2^{\phi\phi}(x,x)H_2^{\phi\phi}(x,y)
-\beta[H_1^h(x)]^2H_2^{\phi\phi}(x,y)+\nonumber \\
&&+\lambda_\phi\left\{3H_3^{\phi\phi\phi}(x,x,y)H_1^\phi(x)
+H_4^{\phi\phi\phi\phi}(x,x,x,y)\right\}= \nonumber \\
&&\beta\left\{
H_4^{hh\phi\phi}(x,x,x,y)+H_3^{hh\phi}(x,x,y)H_1^\phi(x)
+H_2^{hh}(x,x)H_2^{\phi\phi}(x,y)+
2H_3^{h\phi\phi}(x,x,y)H_1^h(x)+\right. \nonumber \\
&&\left.2H_2^{h\phi}(x,x)H_2^{h\phi}(x,y)+
2H_2^{h\phi}(x,y)H_1^\phi(x)
\right\}+\delta^4(x-y),
\eea
and for the Higgs field
\bea
\label{eq:G1h1}
&&\partial^2 H_2^{hh}(x,y)+3\lambda_h[H_1^h(x)]^2H_2^{hh}(x,y)
+3\lambda_h H_2^{hh}(x,x)H_2^{hh}(x,y)
-\beta[H_1^\phi(x)]^2H_2^{hh}(x,y)
\nonumber \\
&&+\lambda_\phi\left\{3H_3^{hhh}(x,x,y)H_1^h(x)
+H_4^{hhhh}(x,x,x,y)\right\}= \nonumber \\
&&\beta\left\{
H_4^{\phi\phi hh}(x,x,x,y)+H_3^{\phi hh}(x,x,y)H_1^h(x)
+H_2^{\phi\phi}(x,x)H_2^{hh}(x,y)+
2H_3^{\phi hh}(x,x,y)H_1^\phi(x)+\right. \nonumber \\
&&\left.2H_2^{\phi h}(x,x)H_2^{\phi h}(x,y)+
2H_2^{\phi h}(x,y)H_1^h(x)\right\}+\delta^4(x-y).
\eea
In order to simplify these equations, we set the 1P-function for the Higgs field to be $H_1^h(x)=v$. This yields
\bea
\label{eq:G1phi1}
&&\partial^2 H_2^{\phi\phi}(x,y)
+3\lambda_\phi [H_1^\phi(x)]^2H_2^{\phi\phi}(x,y)
+3\lambda_\phi H_2^{\phi\phi}(x,x)H_2^{\phi\phi}(x,y)
-\beta v^2H_2^{\phi\phi}(x,y)+\nonumber \\
&&+\lambda_\phi\left\{3H_3^{\phi\phi\phi}(x,x,y)H_1^\phi(x)
+H_4^{\phi\phi\phi\phi}(x,x,x,y)\right\}= \nonumber \\
&&\beta\left\{
H_4^{hh\phi\phi}(x,x,x,y)+H_3^{hh\phi}(x,x,y)H_1^\phi(x)
+H_2^{hh}(x,x)H_2^{\phi\phi}(x,y)+
2vH_3^{h\phi\phi}(x,x,y)+\right. \nonumber \\
&&\left.2H_2^{h\phi}(x,x)H_2^{h\phi}(x,y)+
2H_2^{h\phi}(x,y)H_1^\phi(x)
\right\}+\delta^4(x-y),
\eea
and for the Higgs field
\bea
\label{eq:G1h2}
&&\partial^2 H_2^{hh}(x,y)+3\lambda_hv^2H_2^{hh}(x,y)
+3\lambda_h H_2^{hh}(x,x)H_2^{hh}(x,y)
-\beta[H_1^\phi(x)]^2H_2^{hh}(x,y)
\nonumber \\
&&+\lambda_\phi\left\{3vH_3^{hhh}(x,x,y)
+H_4^{hhhh}(x,x,x,y)\right\}= \nonumber \\
&&\beta\left\{
H_4^{\phi\phi hh}(x,x,x,y)+vH_3^{\phi hh}(x,x,y)
+H_2^{\phi\phi}(x,x)H_2^{hh}(x,y)+
2H_3^{\phi hh}(x,x,y)H_1^\phi(x)+\right. \nonumber \\
&&\left.2H_2^{\phi h}(x,x)H_2^{\phi h}(x,y)+
2vH_2^{\phi h}(x,y)\right\}+\delta^4(x-y).
\eea
These equations can be simplified further if we observe that higher-order nP-functions evaluated at the same space-time points can be chosen to be zero. In this way, one has
\bea
\label{eq:H2phi}
&&\partial^2 H_2^{\phi\phi}(x,y)
+3\lambda_\phi [H_1^\phi(x)]^2H_2^{\phi\phi}(x,y)
+3\lambda_\phi H_2^{\phi\phi}(0)H_2^{\phi\phi}(x,y)
-\beta v^2H_2^{\phi\phi}(x,y)=\nonumber \\
&&\beta\left\{
H_2^{hh}(0)H_2^{\phi\phi}(x,y)+\right. \nonumber \\
&&\left.2H_2^{h\phi}(0)H_2^{h\phi}(x,y)+
2H_2^{h\phi}(x,y)H_1^\phi(x)
\right\}+\delta^4(x-y),
\eea
and for the Higgs field
\bea
\label{eq:H2h}
&&\partial^2 H_2^{hh}(x,y)+3\lambda_hv^2H_2^{hh}(x,y)
+3\lambda_h H_2^{hh}(0)H_2^{hh}(x,y)
-\beta[H_1^\phi(x)]^2H_2^{hh}(x,y)=
\nonumber \\
&&\beta\left\{
H_2^{\phi\phi}(0)H_2^{hh}(x,y)+
2H_2^{\phi h}(0)H_2^{\phi h}(x,y)+
2vH_2^{\phi h}(x,y)\right\}+\delta^4(x-y).
\eea
Therefore, we can introduce the Green functions as
\bea
\label{eq:G2phi}
&&\partial^2 G_2^{\phi\phi}(x,y)
+3\lambda_\phi [H_1^\phi(x)]^2G_2^{\phi\phi}(x,y)
+m_\phi^2G_2^{\phi\phi}(x,y)=\delta^4(x-y) \nonumber \\
&&\partial^2 G_2^{hh}(x,y)+m_h^2G_2^{hh}(x,y)=\delta^4(x-y),
\eea
where we have introduced the mass shift for the $\phi$ field and the mass of the Higgs field as
\bea
\label{eq:masses}
m_\phi^2&=&3\lambda_\phi H_2^{\phi\phi}(0)-\beta v^2 
-\beta H_2^{hh}(0),\nonumber \\
m_h^2&=&3\lambda_hv^2+3\lambda_h H_2^{hh}(0)-\beta H_2^{\phi\phi}(0).
\eea
In the following, we will assume $\beta$ as a small positive parameter. These form a set of two gap equations. Indeed, we can write the following solutions to eqs.(\ref{eq:H2phi}) and (\ref{eq:H2h})
\bea
H_2^{\phi\phi}(x,y)&=&G_2^{\phi\phi}(x,y)+
\beta\int d^4zG_2^{\phi\phi}(x,z)
\left[2H_2^{h\phi}(0)H_2^{h\phi}(z,y)+
2H_2^{h\phi}(z,y)H_1^\phi(z)
\right] \nonumber \\
H_2^{hh}(x,y)&=&G_2^{hh}(x,y)+
\beta\int d^4zG_2^{hh}(x,z)\left[
2H_2^{\phi h}(0)H_2^{\phi h}(z,y)+
2vH_2^{\phi h}(z,y)\right].
\eea
We just note that these are perturbative equations as the cross-correlation functions $H^{h\phi}$ and $H^{\phi h}$ depend on $H^{\phi\phi}$ and $H^{hh}$. The propagators can be written in the form
\be
   G_2^{\phi\phi}(p)=M_\phi{\hat Z}(m_\phi,\lambda_\phi)
   \frac{2\pi^3}{K^3(\kappa)}
	\sum_{n=0}^\infty(-1)^n
	\frac{e^{-(n+\frac{1}{2})\pi\frac{K'(\kappa)}{K(\kappa)}}}{1-e^{-(2n+1)\frac{K'(\kappa)}{K(\kappa)}\pi}}(2n+1)^2\frac{1}{p^2-m_n^2+i\epsilon},
\ee
where
\be
M_\phi=\sqrt{m_\phi^2+\frac{\lambda_\phi\mu^4}{m_\phi^2+\sqrt{m_\phi^4+2\lambda_\phi\mu^4}}},
\ee
${\hat Z}(m_\phi,\lambda_\phi)$ is a given constant, and $\mu$ is an integration constant. The spectrum is given by
\be
\label{eq:ms}
m_n=(2n+1)\frac{\pi}{2K(\kappa)}M_\phi,
\ee
with $K(\kappa)$ being the complete elliptic integral of the first kind.

For the Higgs field, we have
\be
G_2^{hh}(p)=\frac{1}{p^2-m_h^2+i\epsilon}
\ee
as normally used in standard computations. The mass $m_h$ is given in eq.(\ref{eq:masses}) and can be obtained by solving the corresponding set of gap equations. From these results, we can see that the ``phion'' can decay into a number of Higgs particles.

\begin{figure}[H]
\centering
\includegraphics[height=8cm,width=10cm]{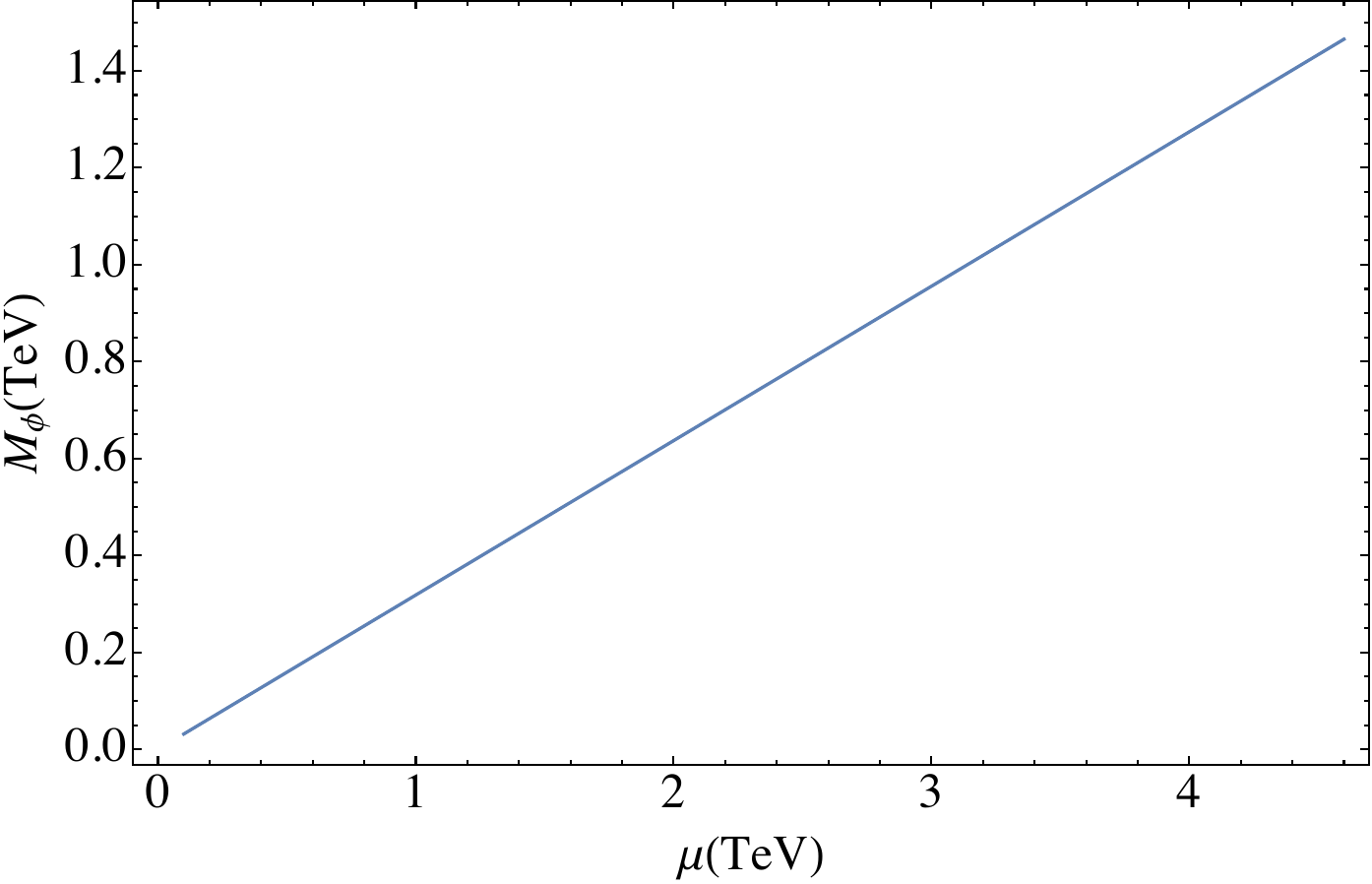}
\includegraphics[height=8cm,width=10cm]{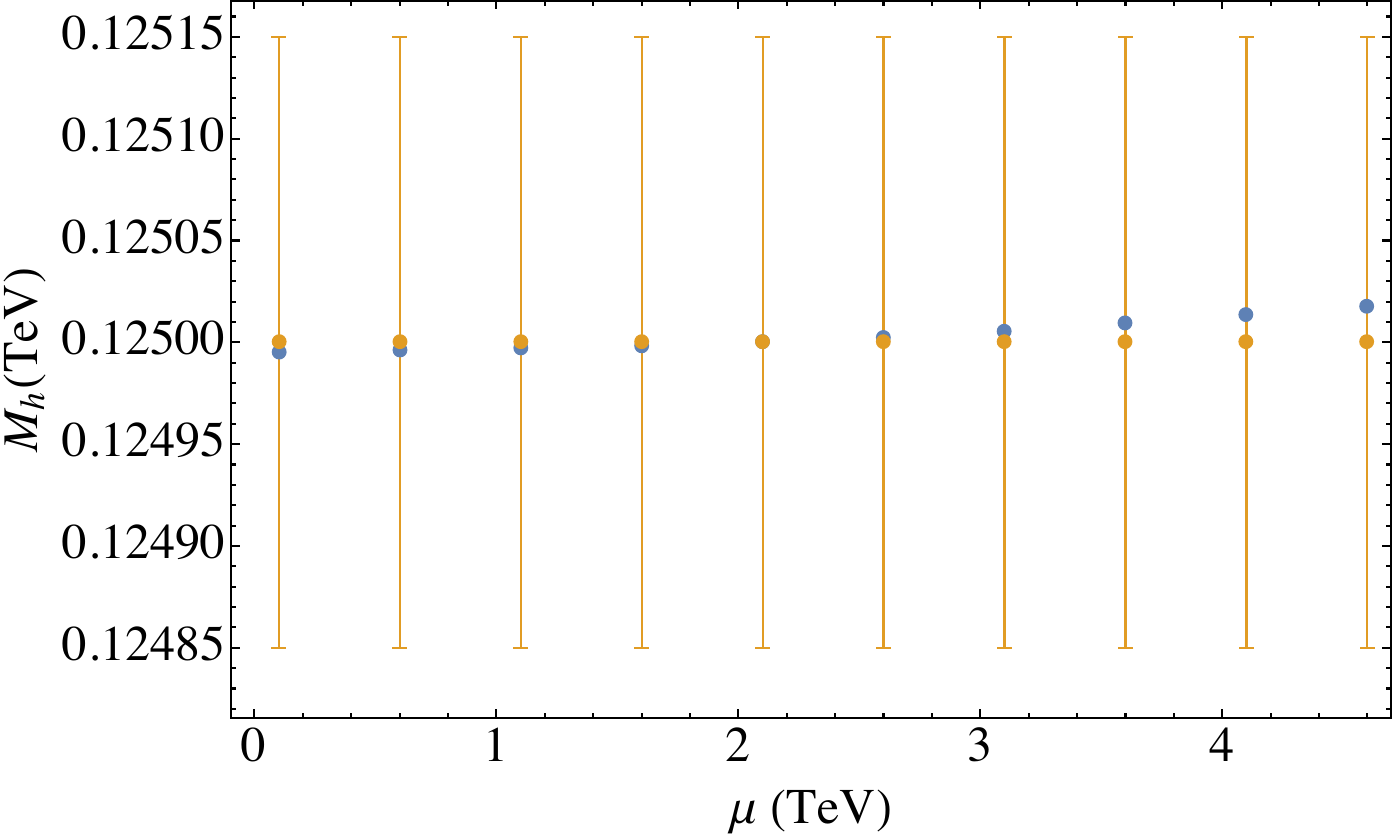}
\caption{\it We show the evolution of the phion mass $M_\phi=m_0$ with respect to $\mu$ and the corresponding evolution of the Higgs boson mass $M_h$ (blue dots) with the value of the experimental Higgs boson mass and its error bar. We set $\beta=10^{-4}$, $\lambda_\phi=10^{-2}$, $\lambda_h=0.086$ and $v=0.246\ \textit{TeV}$. The model appears to be consistent with a wide range of $\mu$ values.}
\label{fig1}
\end{figure}

\begin{figure}[H]
\centering
\includegraphics[height=7cm,width=10cm]{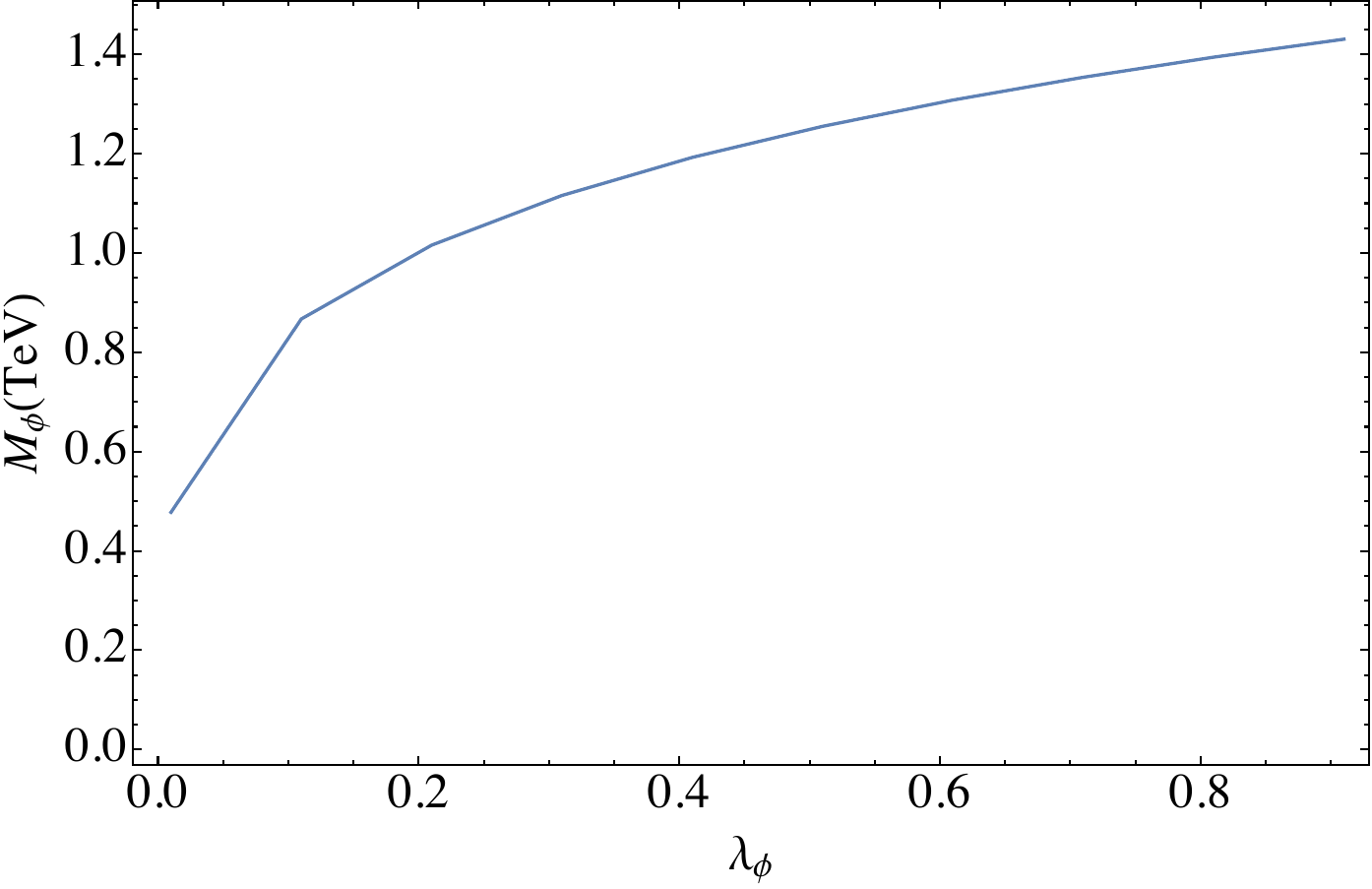}
\includegraphics[height=7cm,width=10cm]{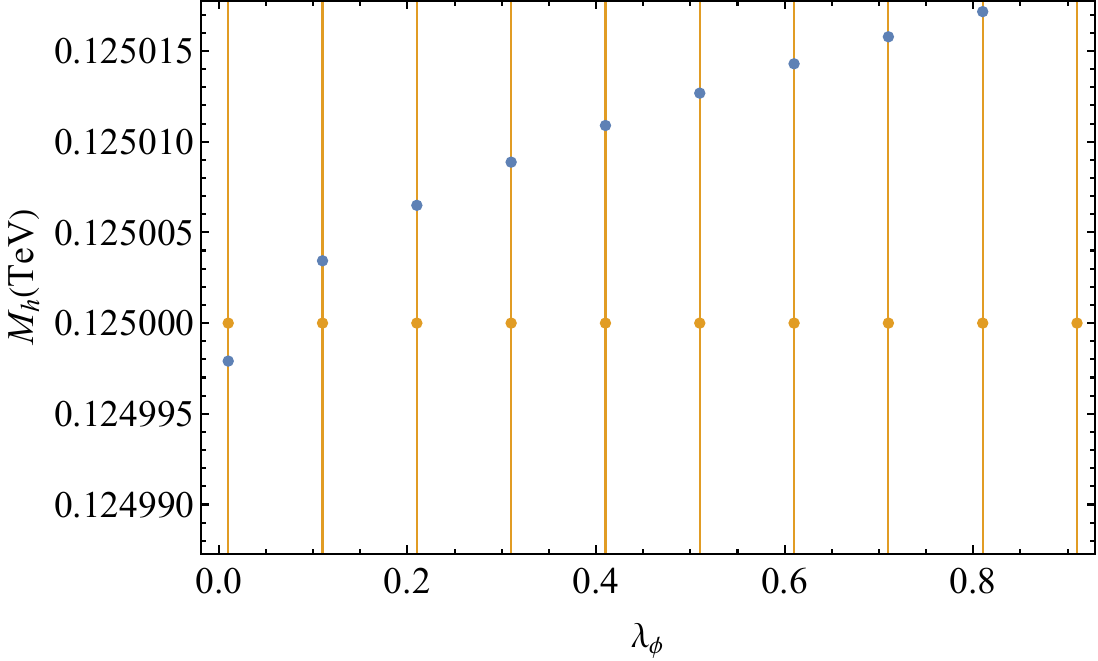}
\caption{\it Plot same as Fig.\ref{fig1} but as a function of $\lambda_\phi$ rather than $\mu$. It is seen that the consistency of the model is granted provided $\lambda_\phi$ is smaller enough. We assume $\mu=1.5\ TeV$.}
\label{fig2}
\end{figure}

At this stage, we can write the gap equations explicitly in the form, keeping Euclidean metric,
\bea
m_\phi^2&=&3\lambda_\phi\int\frac{d^4p}{(2\pi)^4}\sum_n\frac{B_n}{p^2+m_n^2}-\beta v^2 
-\beta\int\frac{d^4p}{(2\pi)^4}\frac{1}{p^2+m_h^2}\nonumber \\
m_h^2&=&3\lambda_hv^2+3\lambda_h\int\frac{d^4p}{(2\pi)^4}\frac{1}{p^2+m_h^2}-\beta \int\frac{d^4p}{(2\pi)^4}\sum_n\frac{B_n}{p^2+m_n^2},
\eea
where
\be
B_n=M_\phi{\hat Z}(m_\phi,\lambda_\phi)
   \frac{2\pi^3}{K^3(\kappa)}(-1)^n
	\frac{e^{-(n+\frac{1}{2})\pi\frac{K'(\kappa)}{K(\kappa)}}}{1-e^{-(2n+1)\frac{K'(\kappa)}{K(\kappa)}\pi}}(2n+1)^2,
\ee
and
\be
m_n(m_\phi)=(2n+1)\frac{\pi}{2K(\kappa)}\sqrt{m_\phi^2+\frac{\lambda_\phi\mu^4}{m_\phi^2+\sqrt{m_\phi^4+2\lambda_\phi\mu^4}}},
\ee
with
\be
\kappa^2 = \frac{m_\phi^2-\sqrt{m_\phi^4+2\lambda_\phi\mu^4}}{m_\phi^2+\sqrt{m_\phi^4+2\lambda_\phi\mu^4}}.
\ee
These equations can be solved very easily if we assume $\beta$ so small to give negligible contributions to these gap equations. For our convenience, we assume $\beta v^2$ finite and retain it. 
Therefore,
\bea
\int\frac{d^4p}{(2\pi)^4}\sum_n\frac{B_n}{p^2+m_n^2}&=&-\frac{1}{16\pi^2}\sum_{n=0}^\infty\frac{\pi^3}{4K^3(i)}(2n+1)^2\frac{e^{-\left(n+\frac{1}{2}\right)\pi}}{1+e^{-(2n+1)\pi}}m_n^2(0), \nonumber \\
\int\frac{d^4p}{(2\pi)^4}\frac{1}{p^2+m_h^2}&=&-\frac{1}{16\pi^2}m_h^2.
\eea
Here, we have assumed the first iterate with the mass shift for the $\phi$ field is zero, and besides, the cut-off terms have been re-absorbed into the coupling constants $\lambda_\phi$ and $\lambda_h$. Therefore, we can finally approximate
\bea
m^2_\phi&=&-\frac{3\lambda_\phi}{256}\frac{\pi^3}{K^5(i)}\sum_{n=0}^\infty(2n+1)^4\frac{e^{-\left(n+\frac{1}{2}\right)\pi}}{1+e^{-(2n+1)\pi}}\sqrt{\frac{\lambda_\phi}{2}}\mu^2-\beta v^2, \nonumber \\
m^2_h&=&3\lambda_h v^2-\frac{3\lambda_h}{16\pi^2}m_h^2.
\eea
We are able to consistently solve this system of equations. E.g., for the Higgs mass one gets
\be
m_h^2=\frac{3\lambda_hv^2}{1+\frac{3\lambda_h}{16\pi^2}}.
\ee
This is consistent with expectations. Taking into consideration the coupling $\beta$, we get the following set of equations
\bea
m^2_\phi&=&-\frac{3\lambda_\phi}{16\pi^2}\sum_{n=0}^\infty\frac{\pi^3}{4K^3(i)}(2n+1)^2\frac{e^{-\left(n+\frac{1}{2}\right)\pi}}{1+e^{-(2n+1)\pi}}m_n^2(0)-\beta v^2+\frac{\beta}{16\pi^2}m_h^2 \nonumber \\
m^2_h&=&3\lambda_h v^2-\frac{3\lambda_h}{16\pi^2}m_h^2+\frac{\beta}{16\pi^2}\sum_{n=0}^\infty\frac{\pi^3}{4K^3(i)}(2n+1)^2\frac{e^{-\left(n+\frac{1}{2}\right)\pi}}{1+e^{-(2n+1)\pi}}m_n^2(0).
\eea
These become
\bea
m^2_\phi&=&-\frac{3\lambda_\phi}{16\pi^2}\sum_{n=0}^\infty\frac{\pi^3}{4K^3(i)}(2n+1)^4\frac{e^{-\left(n+\frac{1}{2}\right)\pi}}{1+e^{-(2n+1)\pi}}m_0^2-\beta v^2+\frac{\beta}{16\pi^2}m_h^2 \nonumber \\
m^2_h&=&3\lambda_h v^2-\frac{3\lambda_h}{16\pi^2}m_h^2+\frac{\beta}{16\pi^2}\sum_{n=0}^\infty\frac{\pi^3}{4K^3(i)}(2n+1)^4\frac{e^{-\left(n+\frac{1}{2}\right)\pi}}{1+e^{-(2n+1)\pi}}m_0^2.
\eea
Here, $m_0$ is the ground state from eq.(\ref{eq:ms}) for the $\phi$ field. Let us introduce a constant,
\be
\xi=\sum_{n=0}^\infty\frac{\pi^3}{4K^3(i)}(2n+1)^4\frac{e^{-\left(n+\frac{1}{2}\right)\pi}}{1+e^{-(2n+1)\pi}}\approx 21.2231\ldots,
\ee
and express the set of equations as
\bea
m^2_\phi&=&-\frac{3\lambda_\phi}{16\pi^2}\xi m_0^2-\beta v^2+\frac{\beta}{16\pi^2}m_h^2, \nonumber \\
m^2_h&=&3\lambda_h v^2-\frac{3\lambda_h}{16\pi^2}m_h^2+\frac{\beta}{16\pi^2}\xi m_0^2.
\eea
The leading order solutions are given above can be written down as
\bea
{\bar m}^2_\phi&=&-\frac{3\lambda_\phi}{16\pi^2}\xi m_0^2-\beta v^2, \nonumber \\
{\bar m}_h^2&=&\frac{3\lambda_hv^2}{1+\frac{3\lambda_h}{16\pi^2}}.
\eea
This yields by iteration
\bea
\label{eq:mphimassh}
m_\phi^2&\approx& {\bar m}^2_\phi+\frac{\beta}{16\pi^2}{\bar m}_h^2, \nonumber \\
M_h^2&\approx& {\bar m}_h^2-\frac{\beta}{3\lambda_\phi}{\bar m}^2_\phi,
\eea
where we neglected $O(\beta^2)$ terms. This implies $\beta/\lambda_\phi\ll 1$ to keep the mass shift for the $h$ field small. With this hypothesis, we can write the phion mass spectrum as
\be
m_n=(2n+1)\frac{\pi}{2K(i)}\left(
\sqrt[4]{\frac{\lambda_\phi}{2}}\mu+\frac{m_\phi^2}{2^\frac{3}{2}\sqrt[4]{\lambda_\phi}\mu}\right).
\ee
The $\mu$ parameter is critical for the physical consistency of the model. In Fig.\ref{fig1} we show how a set of parameters exists that yields a meaningful theoretical result with respect to experimental data.
We also show Fig.\ref{fig2}, which is similar to Fig.\ref{fig1} but as a function of $\lambda_\phi$.
We can achieve consistency.


\begin{figure}[H]
\centering
\includegraphics[width=\textwidth]{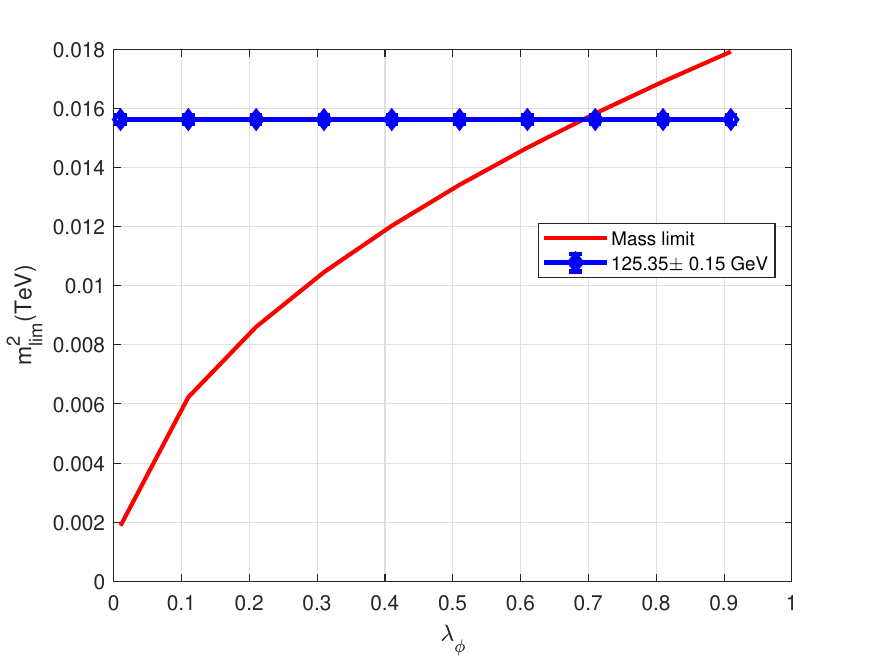}
\caption{\it In this plot, the limit of mass (blue line) is seen to be below the red line for a large set of $\lambda_\phi<1$, in agreement with our discussion in the text. This grants the full consistency of the model. We set $\mu=1.5\ TeV$.}
\label{fig3}
\end{figure}

We can get a general condition for the Higgs vacuum expectation value as it depends on the other parameters of the model. From eq.(\ref{eq:H1h}), we get, setting $H_1^{h}(x)=v$,
\be
\label{eq:ineq}
m_h^2>\frac{\beta}{3\lambda_h}\xi m_0^2.
\ee
We obtain the plot in Fig.\ref{fig3} for the inequality (\ref{eq:ineq}). The red curve is well below the Higgs boson mass for a large set of values of $\lambda_\phi$ as required and so, there exists a meaningful range of parameters for which the scenario is fully consistent as we get the correctly observed EW Higgs mass observed at LHC, being $\lambda_\phi$ fixed (within the uncertainty of Higgs mass data), for $\mu\sim 1.5\ {\rm TeV}$ and the condition (\ref{eq:ineq}) granted.

\medskip




\medskip

\medskip





\medskip

\medskip

\medskip

\section{Conclusions and Discussions}
\label{conc}

We investigated conformally extended Standard Model with a hidden scalar $\phi$ and showed that due to dynamics in the hidden sector with quartic potential (see eq.(\ref{eq:mainmod})), $\phi$ develops a vacuum expectation value (vev) in the form of a mass gap which triggers the electroweak symmetry breaking (EWSB).
We summarise our main findings below:
\begin{itemize}
    \item We provide a novel pathway for dynamical generation of scales, particularly in the context of EW scale generation via dimensional transmutation from a hidden scalar sector starting from a scale-invariant theory at the classical level.
    \item For this purpose we solved the Dyson-Schwinger Equations using the exact solution known via a novel technique developed, by Bender, Milton and Savage \cite{Bender:1999ek}, working in the form of partial differential equations
 (see (\ref{eq:H1phi}),(\ref{eq:H1h1}) and (\ref{eq:G2phi})).
    \item We derived analytically the Higgs boson mass which is dimensionally transmuted from the hidden sector shown in eq.(\ref{eq:mphimassh}).
    \item This yields a consistent solution for the Higgs boson mass, in complete agreement with the experimental data, for a large set of the parameters of the theory for the given ordering. This is very well exemplified in the plots given in Fig.\ref{fig1}, Fig.\ref{fig2} and Fig.\ref{fig3}. 
\end{itemize}
With null signatures of any SM extension at the LHC and in other searches, the framework of naturalness deserves to be re-examined. Among several ideas of explaining the EW scale dynamically generated, we discussed a scenario where conformal symmetry plays an essential
role and the EW scale is a consequence of quantum effects just like QCD scale generation in the SM. We
have shown that it is possible to generate the EW scale by including a new scalar which talks to the SM Higgs via a simple Higgs-portal coupling. The extension is rather minimal. The mass of this new scalar boson is constrained from the successful generation of the SM Higgs boson mass and the BSM microphysics parameters gets fixed within the uncertainties of the Higgs mass measurements. 

There could be a way to search for the Higgs-portal scale in laboratories. Also, our scenario may have an impact on Higgs-portal dark matter model and some profound implications to EW phase transition in the early universe. We leave such investigations to future work.

\section{Acknowledgements}
\label{Asck}

The work of NO is supported in part by the United States Department of Energy (DC-SC 0012447 and DC-SC 0023713).

\section*{Data sharing}

Data sharing not applicable to this article as no datasets were generated or analysed during the current study.

\section*{Conflict of interest}

The authors declare no conflict of interest.


\bibliographystyle{plain} 
\bibliography{ref} 

\begin{thebibliography}{10}

\bibitem{Adler:1982ri}
Stephen~L. Adler.
\newblock {Einstein Gravity as a Symmetry-Breaking Effect in Quantum Field Theory}.
\newblock {\em Rev. Mod. Phys.}, 54:729, 1982.
\newblock [Erratum: Rev.Mod.Phys. 55, 837 (1983)].

\bibitem{AlexanderNunneley:2010nw}
Lisa Alexander-Nunneley and Apostolos Pilaftsis.
\newblock {The Minimal Scale Invariant Extension of the Standard Model}.
\newblock {\em JHEP}, 09:021, 2010.

\bibitem{Altmannshofer:2014vra}
Wolfgang Altmannshofer, William~A. Bardeen, Martin Bauer, Marcela Carena, and Joseph~D. Lykken.
\newblock {Light Dark Matter, Naturalness, and the Radiative Origin of the Electroweak Scale}.
\newblock {\em JHEP}, 01:032, 2015.

\bibitem{1410.1817}
Oleg Antipin, Michele Redi, and Alessandro Strumia.
\newblock {Dynamical generation of the weak and Dark Matter scales from strong interactions}.
\newblock {\em JHEP}, 01:157, 2015.

\bibitem{Baldes:2018emh}
Iason Baldes and Camilo Garcia-Cely.
\newblock {Strong gravitational radiation from a simple dark matter model}.
\newblock {\em JHEP}, 05:190, 2019.

\bibitem{Barman:2021lot}
Basabendu Barman and Anish Ghoshal.
\newblock {Scale invariant FIMP miracle}.
\newblock {\em JCAP}, 03(03):003, 2022.

\bibitem{Barrie:2016rnv}
Neil~D. Barrie, Archil Kobakhidze, and Shelley Liang.
\newblock {Natural Inflation with Hidden Scale Invariance}.
\newblock {\em Phys. Lett. B}, 756:390--393, 2016.

\bibitem{Bender:1999ek}
Carl~M. Bender, Kimball~A. Milton, and Van~M. Savage.
\newblock {Solution of Schwinger-Dyson equations for PT symmetric quantum field theory}.
\newblock {\em Phys. Rev. D}, 62:085001, 2000.

\bibitem{Biswas:2014yia}
Tirthabir Biswas and Nobuchika Okada.
\newblock {Towards LHC physics with nonlocal Standard Model}.
\newblock {\em Nucl. Phys. B}, 898:113--131, 2015.

\bibitem{1807.11490}
Vedran Brdar, Yannick Emonds, Alexander~J. Helmboldt, and Manfred Lindner.
\newblock {Conformal Realization of the Neutrino Option}.
\newblock {\em Phys. Rev. D}, 99(5):055014, 2019.

\bibitem{Brdar:2018num}
Vedran Brdar, Alexander~J. Helmboldt, and Jisuke Kubo.
\newblock {Gravitational Waves from First-Order Phase Transitions: LIGO as a Window to Unexplored Seesaw Scales}.
\newblock {\em JCAP}, 02:021, 2019.

\bibitem{1703.10924}
Ilaria Brivio and Michael Trott.
\newblock {Radiatively Generating the Higgs Potential and Electroweak Scale via the Seesaw Mechanism}.
\newblock {\em Phys. Rev. Lett.}, 119(14):141801, 2017.

\bibitem{1812.01441}
Luca Buoninfante, Anish Ghoshal, Gaetano Lambiase, and Anupam Mazumdar.
\newblock {Transmutation of nonlocal scale in infinite derivative field theories}.
\newblock {\em Phys. Rev. D}, 99(4):044032, 2019.

\bibitem{Callan:1970yg}
Curtis~G. Callan, Jr.
\newblock {Broken scale invariance in scalar field theory}.
\newblock {\em Phys. Rev. D}, 2:1541--1547, 1970.

\bibitem{Chaichian:2018cyv}
Masud Chaichian and Marco Frasca.
\newblock {Condition for confinement in non-Abelian gauge theories}.
\newblock {\em Phys. Lett. B}, 781:33--39, 2018.

\bibitem{Chatterjee:2024dgw}
Arpan Chatterjee, Marco Frasca, Anish Ghoshal, and Stefan Groote.
\newblock {Dynamical generation of electroweak scale from the conformal sector: A strongly coupled Higgs via the Dyson--Schwinger approach}.
\newblock {\em Fortschritte der Physik}, 2025.
\newblock to appear.

\bibitem{Coleman:1970je}
Sidney~R. Coleman and Roman Jackiw.
\newblock {Why dilatation generators do not generate dilatations?}
\newblock {\em Annals Phys.}, 67:552--598, 1971.

\bibitem{Coleman:1973jx}
Sidney~R. Coleman and Erick~J. Weinberg.
\newblock {Radiative Corrections as the Origin of Spontaneous Symmetry Breaking}.
\newblock {\em Phys. Rev. D}, 7:1888--1910, 1973.

\bibitem{Einhorn:2014gfa}
Martin~B. Einhorn and D.~R.~Timothy Jones.
\newblock {Naturalness and Dimensional Transmutation in Classically Scale-Invariant Gravity}.
\newblock {\em JHEP}, 03:047, 2015.

\bibitem{Einhorn:2015lzy}
Martin~B Einhorn and D~R~Timothy Jones.
\newblock {Induced Gravity I: Real Scalar Field}.
\newblock {\em JHEP}, 01:019, 2016.

\bibitem{Einhorn:2016mws}
Martin~B Einhorn and D.~R.~Timothy Jones.
\newblock {Induced Gravity II: Grand Unification}.
\newblock {\em JHEP}, 05:185, 2016.

\bibitem{Englert:2013gz}
Christoph Englert, Joerg Jaeckel, V.~V. Khoze, and Michael Spannowsky.
\newblock {Emergence of the Electroweak Scale through the Higgs Portal}.
\newblock {\em JHEP}, 04:060, 2013.

\bibitem{Farzinnia:2013pga}
Arsham Farzinnia, Hong-Jian He, and Jing Ren.
\newblock {Natural Electroweak Symmetry Breaking from Scale Invariant Higgs Mechanism}.
\newblock {\em Phys. Lett. B}, 727:141--150, 2013.

\bibitem{Farzinnia:2015fka}
Arsham Farzinnia and Seyen Kouwn.
\newblock {Classically scale invariant inflation, supermassive WIMPs, and adimensional gravity}.
\newblock {\em Phys. Rev. D}, 93(6):063528, 2016.

\bibitem{Foot:2007iy}
Robert Foot, Archil Kobakhidze, Kristian~L. McDonald, and Raymond~R. Volkas.
\newblock {A Solution to the hierarchy problem from an almost decoupled hidden sector within a classically scale invariant theory}.
\newblock {\em Phys. Rev. D}, 77:035006, 2008.

\bibitem{Frasca:2005fs}
Marco Frasca.
\newblock {Strong coupling expansion for general relativity}.
\newblock {\em Int. J. Mod. Phys. D}, 15:1373--1386, 2006.

\bibitem{Frasca:2005sx}
Marco Frasca.
\newblock {Strongly coupled quantum field theory}.
\newblock {\em Phys. Rev. D}, 73:027701, 2006.
\newblock [Erratum: Phys.Rev.D 73, 049902 (2006)].

\bibitem{Frasca:2005mv}
Marco Frasca.
\newblock {Dual perturbation expansion for a classical lambda phi**4 field theory}.
\newblock {\em Int. J. Mod. Phys. A}, 22:1441--1450, 2007.

\bibitem{Frasca:2006yx}
Marco Frasca.
\newblock {Proof of triviality of lambda phi**4 theory}.
\newblock {\em Int. J. Mod. Phys. A}, 22:2433--2439, 2007.

\bibitem{Frasca:2007uz}
Marco Frasca.
\newblock {Infrared Gluon and Ghost Propagators}.
\newblock {\em Phys. Lett. B}, 670:73--77, 2008.

\bibitem{Frasca:2008zp}
Marco Frasca.
\newblock {Infrared QCD}.
\newblock {\em Int. J. Mod. Phys. E}, 18:693--703, 2009.

\bibitem{Frasca:2009yp}
Marco Frasca.
\newblock {Mapping a Massless Scalar Field Theory on a Yang-Mills Theory: Classical Case}.
\newblock {\em Mod. Phys. Lett. A}, 24:2425--2432, 2009.

\bibitem{Frasca:2008tg}
Marco Frasca.
\newblock {Yang-Mills Propagators and QCD}.
\newblock {\em Nucl. Phys. B Proc. Suppl.}, 186:260--263, 2009.

\bibitem{Frasca:2010ce}
Marco Frasca.
\newblock {Mapping theorem and Green functions in Yang-Mills theory}.
\newblock {\em PoS}, FACESQCD:039, 2010.

\bibitem{Frasca:2009bc}
Marco Frasca.
\newblock {Exact solutions of classical scalar field equations}.
\newblock {\em J. Nonlin. Math. Phys.}, 18(2):291--297, 2011.

\bibitem{Frasca:2012ne}
Marco Frasca.
\newblock {Classical solutions of a massless Wess-Zumino model}.
\newblock {\em J. Nonlin. Math. Phys.}, 20(4):464--468, 2013.

\bibitem{Frasca:2013tma}
Marco Frasca.
\newblock {Scalar field theory in the strong self-interaction limit}.
\newblock {\em Eur. Phys. J. C}, 74:2929, 2014.

\bibitem{Frasca:2015wva}
Marco Frasca.
\newblock {A theorem on the Higgs sector of the Standard Model}.
\newblock {\em Eur. Phys. J. Plus}, 131(6):199, 2016.

\bibitem{Frasca:2016sky}
Marco Frasca.
\newblock {Confinement in a three-dimensional Yang-Mills theory}.
\newblock {\em Eur. Phys. J. C}, 77(4):255, 2017.

\bibitem{Frasca:2015yva}
Marco Frasca.
\newblock {Quantum Yang-Mills field theory}.
\newblock {\em Eur. Phys. J. Plus}, 132(1):38, 2017.
\newblock [Erratum: Eur.Phys.J.Plus 132, 242 (2017)].

\bibitem{Frasca:2017slg}
Marco Frasca.
\newblock {Spectrum of Yang-Mills theory in 3 and 4 dimensions}.
\newblock {\em Nucl. Part. Phys. Proc.}, 294-296:124--128, 2018.

\bibitem{Frasca:2019ysi}
Marco Frasca.
\newblock {Differential Dyson\textendash{}Schwinger equations for quantum chromodynamics}.
\newblock {\em Eur. Phys. J. C}, 80(8):707, 2020.

\bibitem{Frasca:2020ojd}
Marco Frasca and Anish Ghoshal.
\newblock {Diluted mass gap in strongly coupled non-local Yang-Mills}.
\newblock {\em JHEP}, 21:226, 2020.

\bibitem{2102.10665}
Marco Frasca and Anish Ghoshal.
\newblock {Diluted mass gap in strongly coupled non-local Yang-Mills}.
\newblock {\em JHEP}, 21:226, 2020.

\bibitem{Frasca:2020jbe}
Marco Frasca and Anish Ghoshal.
\newblock {Mass gap in strongly coupled infinite derivative non-local Higgs: Dyson\textendash{}Schwinger approach}.
\newblock {\em Class. Quant. Grav.}, 38(17):175013, 2021.

\bibitem{2011.10586}
Marco Frasca and Anish Ghoshal.
\newblock {Mass gap in strongly coupled infinite derivative non-local Higgs: Dyson\textendash{}Schwinger approach}.
\newblock {\em Class. Quant. Grav.}, 38(17):175013, 2021.

\bibitem{Frasca:2021yuu}
Marco Frasca, Anish Ghoshal, and Stefan Groote.
\newblock {Novel evaluation of the hadronic contribution to the muon\textquoteright{}s g-2 from QCD}.
\newblock {\em Phys. Rev. D}, 104(11):114036, 2021.

\bibitem{Frasca:2021mhi}
Marco Frasca, Anish Ghoshal, and Stefan Groote.
\newblock {Nambu-Jona-Lasinio model correlation functions from QCD}.
\newblock {\em Nucl. Part. Phys. Proc.}, 318-323:138--141, 2022.

\bibitem{Frasca:2022lwp}
Marco Frasca, Anish Ghoshal, and Stefan Groote.
\newblock {Confinement in QCD and generic Yang-Mills theories with matter representations}.
\newblock {\em Phys. Lett. B}, 846:138209, 2023.

\bibitem{Frasca:2022pjf}
Marco Frasca, Anish Ghoshal, and Stefan Groote.
\newblock {Quark confinement in QCD in the 't Hooft limit}.
\newblock {\em Nucl. Part. Phys. Proc.}, 324-329:85--89, 2023.

\bibitem{Frasca:2022duz}
Marco Frasca, Anish Ghoshal, and Alexey~S. Koshelev.
\newblock {Non-perturbative Lee-Wick gauge theory: Towards Confinement \& RGE with strong couplings}.
\newblock {\em Class. Quant. Grav.}, 41(1):015014, 2024.

\bibitem{Frasca:2021iip}
Marco Frasca, Anish Ghoshal, and Nobuchika Okada.
\newblock {Confinement and renormalization group equations in string-inspired nonlocal gauge theories}.
\newblock {\em Phys. Rev. D}, 104(9):096010, 2021.

\bibitem{2201.12267}
Marco Frasca, Anish Ghoshal, and Nobuchika Okada.
\newblock {Fate of false vacuum in non-perturbative regimes}.
\newblock {\em J. Phys. G}, 51(3):035001, 2024.

\bibitem{Frasca:2022kfy}
Marco Frasca, Anish Ghoshal, and Nobuchika Okada.
\newblock {Fate of false vacuum in non-perturbative regimes}.
\newblock {\em J. Phys. G}, 51(3):035001, 2024.

\bibitem{Ghoshal:2018gpq}
Anish Ghoshal.
\newblock {Scalar dark matter probes the scale of nonlocality}.
\newblock {\em Int. J. Mod. Phys. A}, 34(24):1950130, 2019.

\bibitem{1812.02314}
Anish Ghoshal.
\newblock {Scalar dark matter probes the scale of nonlocality}.
\newblock {\em Int. J. Mod. Phys. A}, 34(24):1950130, 2019.

\bibitem{Ghoshal:2017egr}
Anish Ghoshal, Anupam Mazumdar, Nobuchika Okada, and Desmond Villalba.
\newblock {Stability of infinite derivative Abelian Higgs models}.
\newblock {\em Phys. Rev. D}, 97(7):076011, 2018.

\bibitem{1709.09222}
Anish Ghoshal, Anupam Mazumdar, Nobuchika Okada, and Desmond Villalba.
\newblock {Stability of infinite derivative Abelian Higgs models}.
\newblock {\em Phys. Rev. D}, 97(7):076011, 2018.

\bibitem{Ghoshal:2020lfd}
Anish Ghoshal, Anupam Mazumdar, Nobuchika Okada, and Desmond Villalba.
\newblock {Nonlocal non-Abelian gauge theory: Conformal invariance and \ensuremath{\beta}-function}.
\newblock {\em Phys. Rev. D}, 104(1):015003, 2021.

\bibitem{2010.15919}
Anish Ghoshal, Anupam Mazumdar, Nobuchika Okada, and Desmond Villalba.
\newblock {Nonlocal non-Abelian gauge theory: Conformal invariance and \ensuremath{\beta}-function}.
\newblock {\em Phys. Rev. D}, 104(1):015003, 2021.

\bibitem{Ghoshal:2020vud}
Anish Ghoshal and Alberto Salvio.
\newblock {Gravitational waves from fundamental axion dynamics}.
\newblock {\em JHEP}, 12:049, 2020.

\bibitem{Gross:1970tb}
D.~J. Gross and J.~Wess.
\newblock {Scale invariance, conformal invariance, and the high-energy behavior of scattering amplitudes}.
\newblock {\em Phys. Rev. D}, 2:753--764, 1970.

\bibitem{Hambye:2013sna}
Thomas Hambye and Alessandro Strumia.
\newblock {Dynamical generation of the weak and Dark Matter scale}.
\newblock {\em Phys. Rev. D}, 88:055022, 2013.

\bibitem{1306.2329}
Thomas Hambye and Alessandro Strumia.
\newblock {Dynamical generation of the weak and Dark Matter scale}.
\newblock {\em Phys. Rev. D}, 88:055022, 2013.

\bibitem{Hill:2002ap}
Christopher~T. Hill and Elizabeth~H. Simmons.
\newblock {Strong Dynamics and Electroweak Symmetry Breaking}.
\newblock {\em Phys. Rept.}, 381:235--402, 2003.
\newblock [Erratum: Phys.Rept. 390, 553--554 (2004)].

\bibitem{Holthausen:2013ota}
Martin Holthausen, Jisuke Kubo, Kher~Sham Lim, and Manfred Lindner.
\newblock {Electroweak and Conformal Symmetry Breaking by a Strongly Coupled Hidden Sector}.
\newblock {\em JHEP}, 12:076, 2013.

\bibitem{0902.4050}
Satoshi Iso, Nobuchika Okada, and Yuta Orikasa.
\newblock {Classically conformal $B^-$ L extended Standard Model}.
\newblock {\em Phys. Lett. B}, 676:81--87, 2009.

\bibitem{0909.0128}
Satoshi Iso, Nobuchika Okada, and Yuta Orikasa.
\newblock {The minimal B-L model naturally realized at TeV scale}.
\newblock {\em Phys. Rev. D}, 80:115007, 2009.

\bibitem{1210.2848}
Satoshi Iso and Yuta Orikasa.
\newblock {TeV Scale B-L model with a flat Higgs potential at the Planck scale: In view of the hierarchy problem}.
\newblock {\em PTEP}, 2013:023B08, 2013.

\bibitem{Iso:2017uuu}
Satoshi Iso, Pasquale~D. Serpico, and Kengo Shimada.
\newblock {QCD-Electroweak First-Order Phase Transition in a Supercooled Universe}.
\newblock {\em Phys. Rev. Lett.}, 119(14):141301, 2017.

\bibitem{Jaeckel:2016jlh}
Joerg Jaeckel, Valentin~V. Khoze, and Michael Spannowsky.
\newblock {Hearing the signal of dark sectors with gravitational wave detectors}.
\newblock {\em Phys. Rev. D}, 94(10):103519, 2016.

\bibitem{Kannike:2015fom}
Kristjan Kannike, G.~H\"utsi, L.~Pizza, A.~Racioppi, M.~Raidal, A.~Salvio, and Alessandro Strumia.
\newblock {Dynamically Induced Planck Scale and Inflation}.
\newblock {\em PoS}, EPS-HEP2015:379, 2015.

\bibitem{Kannike:2015apa}
Kristjan Kannike, Gert H\"utsi, Liberato Pizza, Antonio Racioppi, Martti Raidal, Alberto Salvio, and Alessandro Strumia.
\newblock {Dynamically Induced Planck Scale and Inflation}.
\newblock {\em JHEP}, 05:065, 2015.

\bibitem{Kannike:2016bny}
Kristjan Kannike, Giulio~Maria Pelaggi, Alberto Salvio, and Alessandro Strumia.
\newblock {The Higgs of the Higgs and the diphoton channel}.
\newblock {\em JHEP}, 07:101, 2016.

\bibitem{Kannike:2014mia}
Kristjan Kannike, Antonio Racioppi, and Martti Raidal.
\newblock {Embedding inflation into the Standard Model - more evidence for classical scale invariance}.
\newblock {\em JHEP}, 06:154, 2014.

\bibitem{Karam:2015jta}
Alexandros Karam and Kyriakos Tamvakis.
\newblock {Dark matter and neutrino masses from a scale-invariant multi-Higgs portal}.
\newblock {\em Phys. Rev. D}, 92(7):075010, 2015.

\bibitem{Karam:2016rsz}
Alexandros Karam and Kyriakos Tamvakis.
\newblock {Dark Matter from a Classically Scale-Invariant $SU(3)_X$}.
\newblock {\em Phys. Rev. D}, 94(5):055004, 2016.

\bibitem{Khoze:2013uia}
Valentin~V. Khoze.
\newblock {Inflation and Dark Matter in the Higgs Portal of Classically Scale Invariant Standard Model}.
\newblock {\em JHEP}, 11:215, 2013.

\bibitem{Kilic:2009mi}
Can Kilic, Takemichi Okui, and Raman Sundrum.
\newblock {Vectorlike Confinement at the LHC}.
\newblock {\em JHEP}, 02:018, 2010.

\bibitem{Klett:2022iga}
Sophie Klett, Manfred Lindner, and Andreas Trautner.
\newblock {Generating the electro-weak scale by vector-like quark condensation}.
\newblock {\em SciPost Phys.}, 14(4):076, 2023.

\bibitem{Lane:2002sm}
Kenneth Lane and Stephen Mrenna.
\newblock {The Collider Phenomenology of Technihadrons in the Technicolor Straw Man Model}.
\newblock {\em Phys. Rev. D}, 67:115011, 2003.

\bibitem{Marzo:2018nov}
Carlo Marzo, Luca Marzola, and Ville Vaskonen.
\newblock {Phase transition and vacuum stability in the classically conformal B\textendash{}L model}.
\newblock {\em Eur. Phys. J. C}, 79(7):601, 2019.

\bibitem{Marzola:2017jzl}
Luca Marzola, Antonio Racioppi, and Ville Vaskonen.
\newblock {Phase transition and gravitational wave phenomenology of scalar conformal extensions of the Standard Model}.
\newblock {\em Eur. Phys. J. C}, 77(7):484, 2017.

\bibitem{Prokopec:2018tnq}
Tomislav Prokopec, Jonas Rezacek, and Bogumi\l{}a \'Swie\.zewska.
\newblock {Gravitational waves from conformal symmetry breaking}.
\newblock {\em JCAP}, 02:009, 2019.

\bibitem{Rinaldi:2014gha}
Massimiliano Rinaldi, Guido Cognola, Luciano Vanzo, and Sergio Zerbini.
\newblock {Inflation in scale-invariant theories of gravity}.
\newblock {\em Phys. Rev. D}, 91(12):123527, 2015.

\bibitem{2012.11608}
Alberto Salvio.
\newblock {Dimensional Transmutation in Gravity and Cosmology}.
\newblock {\em Int. J. Mod. Phys. A}, 36(08n09):2130006, 2021.

\bibitem{Salvio:2014soa}
Alberto Salvio and Alessandro Strumia.
\newblock {Agravity}.
\newblock {\em JHEP}, 06:080, 2014.

\bibitem{Sannino:2008ha}
Francesco Sannino.
\newblock {Dynamical Stabilization of the Fermi Scale: Phase Diagram of Strongly Coupled Theories for (Minimal) Walking Technicolor and Unparticles}.
\newblock 4 2008.

\bibitem{Su:2021qvm}
Xing-Fu Su, You-Ying Li, Rosy Nicolaidou, Min Chen, Hsin-Yeh Wu, and Stathes Paganis.
\newblock {High $P_T$ Higgs excess as a signal of non-local QFT at the LHC}.
\newblock {\em Eur. Phys. J. C}, 81(9):796, 2021.

\bibitem{Tambalo:2016eqr}
Giovanni Tambalo and Massimiliano Rinaldi.
\newblock {Inflation and reheating in scale-invariant scalar-tensor gravity}.
\newblock {\em Gen. Rel. Grav.}, 49(4):52, 2017.

\end{thebibliography}
\end{document}